# *Ab initio* calculations of the atomic and electronic structures of crystalline PEO$_3$:LiCF$_3$SO$_3$ electrolytes

Sha Xue[a,d], Yingdi Liu[b], Hongli Dang[b], Yaping Li[b], Dale Teeters[c,e], Daniel W. Crunkleton[a,d,e], and Sanwu Wang[b,d,e,*]

[a] Russell School of Chemical Engineering, The University of Tulsa, Tulsa, OK 74104, USA

[b] Department of Physics and Engineering Physics, The University of Tulsa, Tulsa, OK 74104, USA

[c] Department of Chemistry and Biochemistry, The University of Tulsa, Tulsa, OK 74104, USA

[d] Tulsa Institute of Alternative Energy, The University of Tulsa, Tulsa, OK 74104, USA

[e] Tulsa Institute of Nanotechnology, The University of Tulsa, Tulsa, OK 74104, USA

**Abstract**

The PEO$_3$:LiCF$_3$SO$_3$ polymer electrolyte has attracted significant research due to its enhanced stability at the lithium/polymer interface of high conductivity polymer batteries. Experimental studies have shown that, depending on the preparation conditions, both the PEO$_3$:LiCF$_3$SO$_3$ crystalline complex and the PEO$_3$:LiCF$_3$SO$_3$ amorphous phase can be formed. However, previous theoretical investigations focused on the short chain amorphous PEO$_3$:LiCF$_3$SO$_3$ system. We report *ab initio* density-functional-theory calculations of crystalline PEO$_3$:LiCF$_3$SO$_3$. The calculated results about the bonding configuration, electronic structures, and conductivity properties are in good agreement with the experimental measurements.

**Keywords**: *Ab initio* calculations, Polymer electrolytes, Inorganic lithium salts, Ionic conductivity, Crystal structures, Electronic structures.

* Corresponding author. Tel.: +1 918 631 3022. E-mail address: sanwu-wang@utulsa.edu (S. Wang).

# 1. Introduction

With the advent of high conductivity polymer batteries, a great deal of research has been performed on gel polymer electrolytes, in which a liquid plasticizer is entrapped in a polymer matrix [1]. These polymer electrolytes are generally composites of a poly (poly ethylene oxide) or another modified polymer and a salt, such as $LiCF_3SO_3$, $Li(CF_3SO_2)_2N$, $LiClO_4$, $LiPF_6$, or $LiAsF_6$. Such electrolytes, however, have technological issues that prevent them from being fully developed. For example, gel electrolyte systems are difficult to use with lithium anodes because of their instability with lithium [2-4]. As a result, polymer electrolytes that contain no plasticizers or solutions, *i.e.* "dry" polymer electrolytes, are preferable because of their enhanced stability at the lithium/polymer interface [5]. Specific research attention has been placed on PEO polymers with the general formula $(-CH_2-CH_2-O-)_n$ and which form complexes with inorganic lithium salts. In particular, $PEO_3$: $LiCF_3SO_3$ is one of the most investigated systems.

Numerous experimental and theoretical studies have been conducted on the $PEO_3$: $LiCF_3SO_3$ electrolyte system [5-18]. Experimental research has demonstrated that, depending on the preparation conditions including particularly the temperature, both crystalline and amorphous phases of $PEO_3$: $LiCF_3SO_3$ can be formed [5-18]. The difference in conductivity between the crystalline and amorphous phases has been also studied experimentally [6-8]. Some authors conclude that the conductivity in the crystalline phases is much lower than that in the amorphous phase [6]. Other studies have made the contrary conclusion, namely, ionic conductivity in the crystalline phase can actually be greater than in the amorphous materials when the temperature exceeds the glass transition temperature, $T_g$ [7,8]. These seemingly inconsistent conclusions concerning the influence of the phase on the conductivity suggest that more fundamental investigations of the ionic conductivity process in $PEO_3$:$LiCF_3SO_3$ electrolytes are needed.

The atomic structure of crystalline $PEO_3$:$LiCF_3SO_3$ was determined by both powder and single crystal X-ray diffraction (XRD) measurements [5,14]. Theoretical investigations for the local structure of short chain $PEO_3$:$LiCF_3SO_3$ were also performed [9-11]. As these studies employed cluster models, which do not exhibit the periodic nature of crystalline $PEO_3$:$LiCF_3SO_3$, theoretical understanding of the *crystalline* $PEO_3$:$LiCF_3SO_3$ phase at the atomic-scale is still lacking. Furthermore, characterization of the ionic conductivity of $PEO_3$:$LiCF_3SO_3$ requires a fundamental understanding of both atomic and electronic structures. In this study, we use *ab initio* density-functional theory (DFT) to determine the atomic structure and electronic properties of crystalline $PEO_3$:$LiCF_3SO_3$. The obtained results based on the DFT calculations are also compared with the available experimental data.

The remainder of this paper is as follows. In Section 2, we provide details about the computational method and the supercell models that were used. In Section 3, we present and discuss the theoretical results including comparison with the experimental data. Finally, in Section 4, we summarize the conclusions obtained from the DFT calculations.

## 2. Model and Method

The crystal structure of $PEO_3$:$LiCF_3SO_3$ has been determined previously by powder X-ray diffraction [5]. The atomic structure of $PEO_3$:$LiCF_3SO_3$ was also determined by single crystal XRD measurements [14]. The structural parameters obtained from both powder and single crystal XRD measurements are essential the same (with some differences, as discussed in Section 3). As the structural data obtained from powder XRD is thorough, it was adopted to construct the initial model for our DFT calculations. The unit cell is experimentally determined to be monoclinic, to have a space group P21/a, and with lattice parameters $a$ = 16.768 Å, $b$ = 8.613 Å, $c$ = 10.070 Å, and $\beta$ = 121.02 °. The PEO chains adopt a helical conformation with all C-O bond T (Trans) and C-C bonds either G (gauche) or G- (gauche minus). Three ethylene oxide units are involved in the basic repeating sequence which is TTGTTGTTG-. The cation is coordinated by three ether oxygen atoms from the chain and one oxygen from each of two $CF_3SO_3^-$ groups completing the five-coordinated environment. There is no ionic cross-linking between chains, which are connected through weak van der Waals interactions [13].

Fig. 1 shows the initial model used in the present investigation. The model was based on the refined atomic parameters (shown in Table 1) for $PEO_3:LiCF_3SO_3$ at 298 K [5]. The unit cell of the model contains 28 C atoms, 48 H atoms, 24 O atoms, 4 Li atoms, 4 S atoms, and 12 F atoms. Periodic boundary conditions were used in the calculations.

The *ab initio* calculations are based on density-functional theory (DFT), the projector augmented wave (PAW) method, and plane-wave basis sets [19-26]. The outmost cutoff radius, the partial core radius, and the radius of the PAW sphere for the PAW potential of C were 1.500 Å, 1.200 Å, and 1.501 Å, respectively. The outmost cutoff radius for O (Li, S, F, and H) was 1.520 Å (2.050 Å, 1.900 Å, 1.520 Å, and 1.100 Å); the partial core radius for O (Li, S, F, and H) was 1.200 Å (1.500 Å, 1.500 Å, 1.200 Å, and 0.000 Å); and the radius of the PAW sphere for O (Li, S, F, and H) was 1.550 Å (2.094 Å, 1.954 Å, 1.539 Å, and 1.112 Å), respectively [23,24]. The electronic configurations of C, O, Li, S, F, and H were $[He]2s^22p^2$, $[He]2s^22p^4$, $[He]2s^1$, $[Ne]3s^23p^4$, $[He]2s^22p^5$, and $1s^1$, respectively. The results reported in this paper were obtained using the Vienna *ab initio* simulation package (VASP) [20-22]. The exchange-correlation effects were treated with the Perdew-Burke-Ernzerhof (PBE) scheme [25], with an energy cut-off of 400 eV and 2 special **k** points in the Brillouin zone of the supercell for all calculations. All the structures were fully relaxed until the forces on all atoms were smaller than 0.01 eV/ Å.

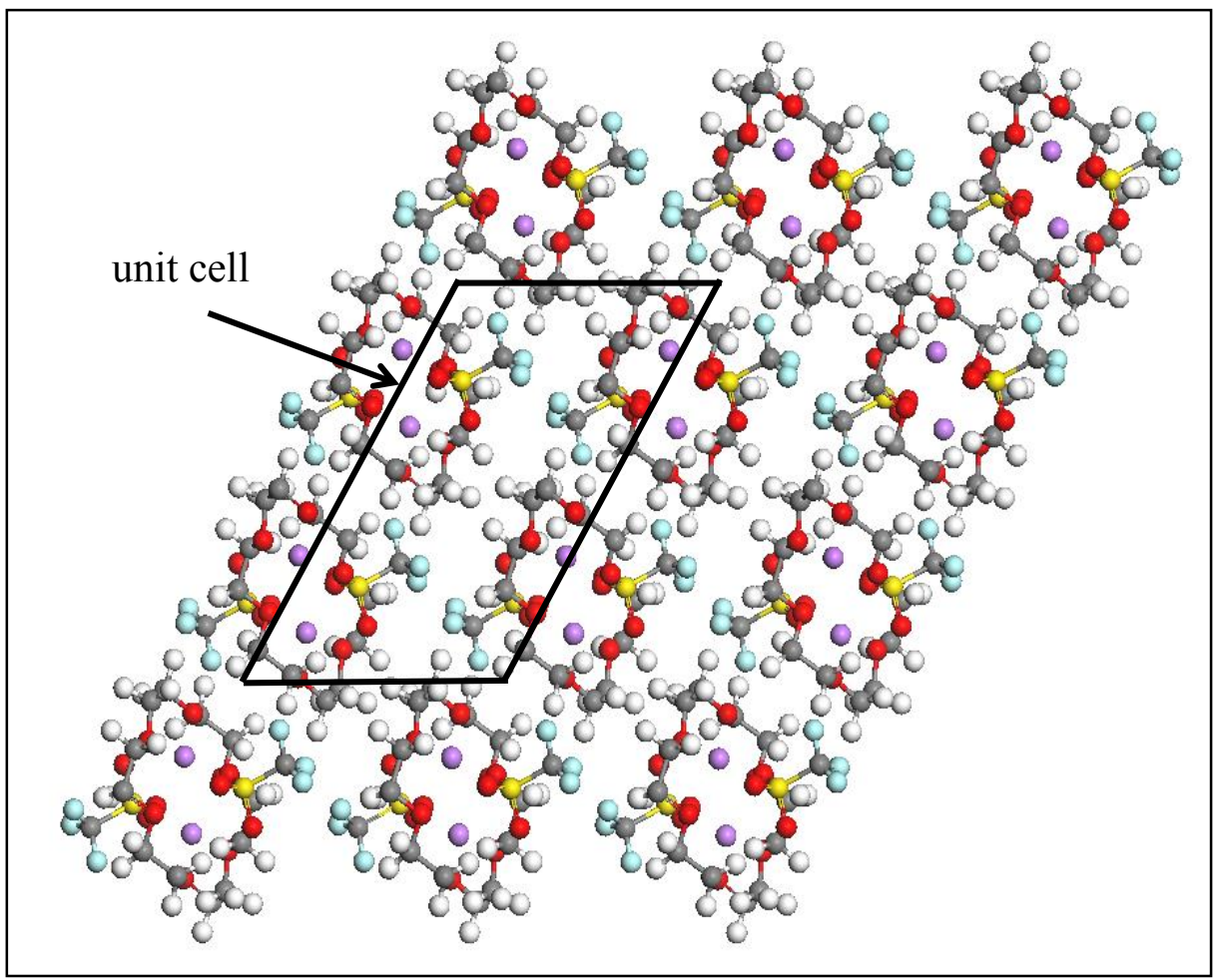

Fig. 1 (Color online). Structure along the fiber axis of experimental crystalline $PEO_3$:$LiCF_3SO_3$. The structure is periodically repeated. The unit cell contains 28 carbon atoms (grey), 48 hydrogen atoms (white), 24 oxygen atoms (red), 4 lithium atoms (purple), 4 sulfur atoms (yellow), and 12 fluorine atom (light blue). Note that the carbon repeating unit is TTGTTGTTG$^-$.

Table 1. Refined atomic parameters for PEO$_3$:LiCF$_3$SO$_3$ at 298 K (fractional coordinates) are taken from Ref. [5] as the initial geometry for *ab initio* calculations. Hydrogen atoms were included in the refinement but are not reported.

| Atom | x | y | z |
| --- | --- | --- | --- |
| O1 | 0.206 | 0.110 | 0.090 |
| C1 | 0.262 | 0.091 | 0.253 |
| C2 | 0.366 | 0.099 | 0.298 |
| O2 | 0.386 | 0.246 | 0.265 |
| C3 | 0.486 | 0.261 | 0.341 |
| C4 | 0.513 | 0.426 | 0.314 |
| O3 | 0.455 | 0.469 | 0.158 |
| C5 | 0.450 | 0.631 | 0.131 |
| C6 | 0.387 | 0.651 | −0.051 |
| S1 | 0.270 | 0.124 | 0.809 |
| C7 | 0.328 | 0.077 | 0.703 |
| O4 | 0.277 | −0.001 | 0.903 |
| O5 | 0.303 | 0.264 | 0.893 |
| O6 | 0.172 | 0.150 | 0.693 |
| F1 | 0.321 | 0.204 | 0.624 |
| F2 | 0.284 | −0.031 | 0.601 |
| F3 | 0.413 | 0.041 | 0.798 |
| Li1 | 0.156 | 0.903 | 0.906 |

# 3. Results and Discussion

Lattice constants were optimized and the optimized values of the lattice parameters are shown in Table 2. We started with the experimental parameters and calculated the total energies of the structures as functions of the three lattice constants with a fixed value of the angle $\beta$ (three lattice constants vary with the same rate in order to keep the angle $\beta$ fixed), and we then changed the value of $\beta$ and calculated the total energies as functions of the lattice constants again. In each calculation, the atomic structure was relaxed. The optimized lattice constants were taken as those of the minimum-energy structure. Our DFT calculations yielded the lattice constants of $a$ = 17.268 Å, $b$ = 8.870 Å, and $c$ = 10.370 Å, each of which is only slightly (~ 3.0%) greater than the corresponding experimental value [5]. The optimized value of the angle $\beta$, 121.04 °, is practically the same as experimentally measured value of 121.02 °[5].

Table 2. Lattice constants of crystalline $PEO_3$:$LiCF_3SO_3$ determined by *ab initio* DFT calculations. Experimental data are also shown for comparison.

|  | $a$ (Å) | $b$ (Å) | $c$ (Å) | $\beta$ (°) |
| --- | --- | --- | --- | --- |
| This work | 17.268 | 8.870 | 10.370 | 121.04 |
| Experiment[a] | 16.768 | 8.613 | 10.070 | 121.02 |
| Difference (%) | 2.98 | 2.98 | 2.98 | 0.02 |

[a]Ref. [5]

Fig. 2 illustrates the structure of the crystalline PEO$_3$:LiCF$_3$SO$_3$ obtained from the relaxation calculation with the optimized lattice parameters. The relaxed structure, as shown in Fig. 2, is observed to have the same bonding configuration as the experimental structure (Fig. 1). In particular, both the experimental and relaxed structures have lithium atom maintaining threefold coordination with ether oxygen atoms from the PEO chain and twofold coordination with oxygen atoms from each of two nearby CF$_3$SO$_3^-$ groups (Fig. 3). Note additionally that the relaxed and experimental structures both contain the repeating conformation sequence of TTGTTGTTG$^-$. However, the relaxed structure shows variations in inter-atomic distances and angles (see discussions below).

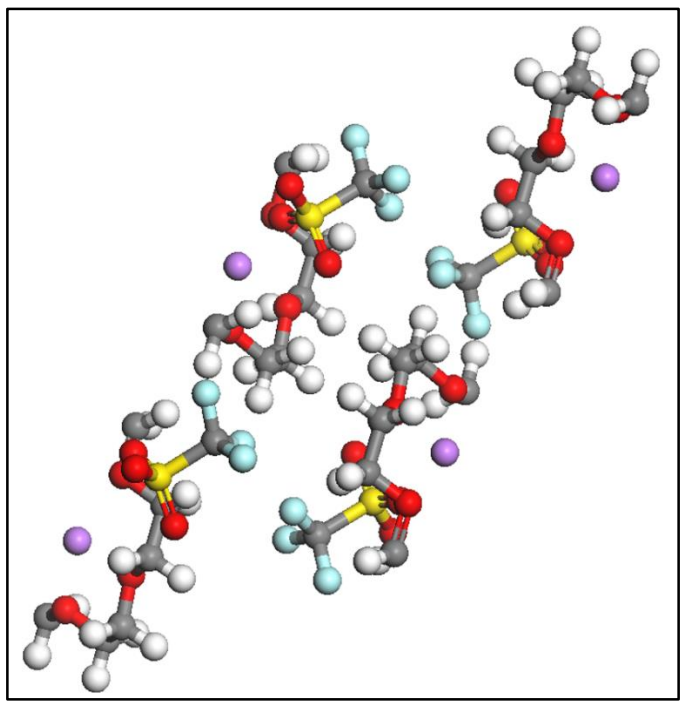

Fig. 2 (Color online). Schematic of the structure of crystalline PEO$_3$:LiCF$_3$SO$_3$ obtained after relaxation. Note that the TTGTTGTTG- sequence is maintained as in the experimentally determined structure.

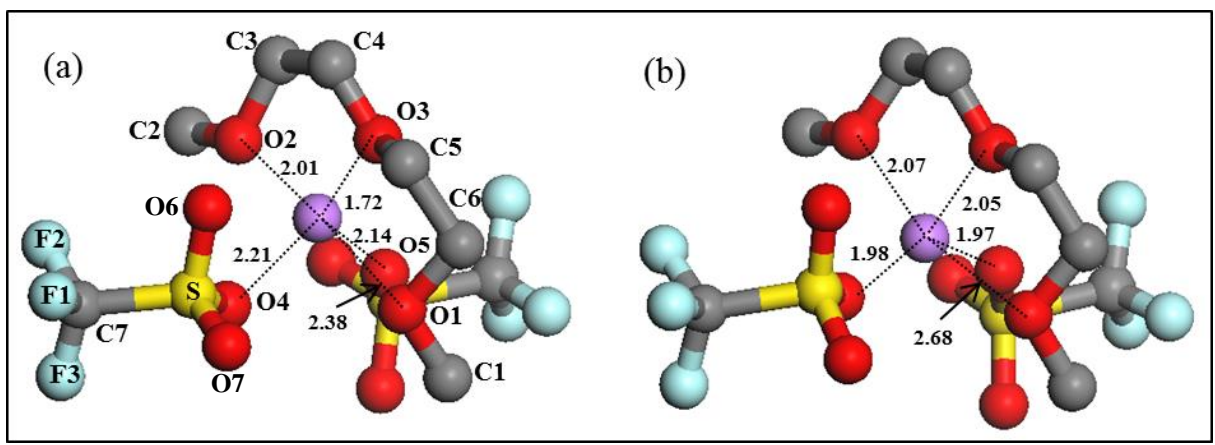

Fig. 3 (Color online). Schematics of the local structure containing a Li atom determined by (a) powder X-ray diffraction (Ref. [5]); and (b) *Ab initio* density-functional theory (this work). Distances between lithium and its nearby five oxygen atoms are also shown.

Tables 3 and 4 present the distances and angles between the different atoms, and some differences between the calculated values and the experimental values obtained with powder XRD are observed. More specifically, as shown in Table 3, the distances between the ether oxygens and lithium for the calculated values are all greater than the experimental values, and the distances of Li1-O1 and Li1-O3 (ether oxygens located in the sides) are larger by more than 10%. On the other hand, however, the distances of Li1-O4 and Li1-O5 are lower by about 12% and 9%, respectively. These results suggest that the computed position of the lithium atom has moved slightly from the oxygen atoms on the PEO chain to the oxygen atoms of the $CF_3SO_3^-$ groups. The structure determined with single crystal XRD shows that the distances of Li1-O1, Li1-O2 and Li1-O3 are 2.46, 2.07 and 2.05 Å, respectively, as shown in Table 3 (in parentheses) [14]. The corresponding theoretical distances (2.68, 2.08, and 2.05 Å) are thus in better agreement with the single crystal XRD (some difference may be due to the fact that the single crystal sample contained defects). In addition, the experimental work of single-crystal XRD showed that O1 is directed toward a second lithium (Li2) with a distance of 3.05 Å (not shown in Table 3), indicating an additional very weak coordination bond [14]. The corresponding distance determined with powder XRD is 3.41 Å [5]. Our theoretical distance of Li2-O1 is 3.01 Å, in good agreement with the single-crystal XRD data (note that the numbers for the oxygen atoms used in Ref. [14] are different to what we used in this work).

Table 3. Comparison of the bond lengths between the experimental data by powder XRD (Ref. [5]) and the *ab initio* results (this work). Available bond lengths determined by single crystal XRD [Ref. 14] are also shown (in parentheses) for comparison. Note that other bond lengths in the table are not available from single crystal XRD.

| Atom-Atom | Experimental Distance (Å) | *Ab Initio* Distance (Å) | Difference (%) |
|---|---|---|---|
| O1-C1 | 1.42 | 1.43 | 0.70 |
| C1-C2 | 1.56 | 1.51 | −3.31 |
| C2-O2 | 1.40 | 1.43 | 2.10 |
| O2-C3 | 1.45 | 1.43 | −1.40 |
| C3-C4 | 1.55 | 1.51 | −2.65 |
| C4-O3 | 1.41 | 1.43 | 1.40 |
| O3-C5 | 1.41 | 1.44 | 2.08 |
| C5-C6 | 1.58 | 1.51 | −4.64 |
| C6-O1 | 1.44 | 1.43 | −0.70 |
| S1-O4 | 1.39 | 1.46 | 4.79 |
| S1-O5 | 1.42 | 1.46 | 2.74 |
| S1-O6 | 1.46 | 1.45 | −0.69 |
| S1-C7 | 1.83 | 1.83 | 0.00 |
| C7-F1 | 1.31 | 1.36 | 3.68 |
| C7-F2 | 1.30 | 1.36 | 4.41 |
| C7-F3 | 1.28 | 1.36 | 5.88 |
| Li1-O1 | 2.38 (2.47) | 2.68 | 11.19 (7.84) |
| Li1-O2 | 2.01 (2.08) | 2.07 | 2.90 (−0.48) |
| Li1-O3 | 1.72 (2.05) | 2.05 | 16.10 (0.00) |
| Li1-O4 | 2.21 | 1.98 | −11.62 |
| Li1-O5 | 2.14 | 1.97 | −8.63 |

Additionally, as shown in Table 4, the computed angles of O1-Li1-O2, O1-Li1-O3 and O2-Li1-O3 are all lower by more than 13%, consistent with the conclusion of the lithium atom moving away slightly from O1, O2 and O3. As the lithium atom moves closer to O4 and O5, the angle of O4-Li1-O5 is observed to increase from 112° to 135°. Table 4 also shows that the angle of O2-Li1-O5 increases from 104° to 109° while the angle of O3-Li1-O4 is reduced from 134° to 117°. This indicates that the Li atom moves even closer to O5 than O4 in order to reach its most stable configuration.

In order to understand the electronic properties, we present calculated total electronic density of states (DOS) of crystalline $PEO_3$:$LiCF_3SO_3$ and the DOS projected onto individual components. Fig. 4 shows the total and projected DOS for crystalline $PEO_3$:$LiCF_3SO_3$. For comparison, the total DOS for isolated $LiCF_3SO_3$ is also shown in Fig. 4(d).

A significant feature of the electronic structure is its wide band gap, calculated to be approximately 4.3 eV. Such a band gap suggests the absence of electronic conductivity in $PEO_3$:$LiCF_3SO_3$, consistent with experimental conclusions of the predominance of Li ion conductivity. The detailed mechanism for the ionic conductivity for both crystalline and amorphous PEO$_3$:LiCF$_3$SO$_3$ at the atomic scale still remains elusive, however. Measurements with the pulsed magnetic field gradient technique for a related solid polymer electrolyte, (PEO)$_n$:Li(CF$_3$SO$_2$)$_2$N, showed that ion transport of both cations (Li$^+$) and anions (F$^-$-containing species) in both crystalline and amorphous phases made contribution to ion conductivity [27]. Meanwhile, the activation energies for the cationic diffusion were determined to be smaller than those for the anionic diffusion. However, the experimental measurements could not be used to determine the diffusion pathways. On the other hand, NMR studies suggested that ionic conductivity in the crystalline polymer electrolytes was dominated by the transport of lithium ions while both cations and anions were found to contribute to conductivity in the amorphous phases [7]. Palma *et al.* used *ab initio* molecular dynamics simulations to determine the diffusion of $Li^+$ in the crystalline PEO chain, but anions were not included [28]. On the other hand, cluster models were employed by Johansson *et al.* [11,29] and by Yarmolenko *et al.* [30] to study diffusion of $Li^+$ in PEO with and without the presence of a Li salt ($LiPF_6$ or $LiClO_4$), but periodicity of PEO was not included in the calculations. While diffusion pathways and activation energies of ions in $PEO_3$:$LiCF_3SO_3$ have not been determined with *ab initio* calculations, the present work for the structure with the use of the supercell model, which includes long-range order of PEO and both cations and anions, provides an excellent start point for the future study on ion transport.

Table 4. Comparison of the bond angles between the experimental data (Ref. [5]) and *ab initio* results (this work).

| Atoms | Experimental Angle (°) | Ab Initio Angle (°) | Difference (%) |
|---|---|---|---|
| C6-O1-C1 | 108 | 111 | 2.70 |
| C2-O2-C3 | 109 | 113 | 3.54 |
| O2-C3-C4 | 111 | 107 | −3.74 |
| C3-C4-O3 | 110 | 107 | −2.80 |
| C4-O3-C5 | 115 | 114 | −0.88 |
| O3-C5-C6 | 105 | 108 | 2.78 |
| C5-C6-O1 | 106 | 109 | 2.75 |
| O1-Li1-O2 | 174 | 151 | −15.23 |
| O1-Li1-O3 | 85 | 72 | −18.06 |
| O1-Li1-O4 | 77 | 82 | 6.10 |
| O1-Li1-O5 | 83 | 89 | 6.74 |
| O2-Li1-O3 | 94 | 82 | −14.63 |
| O2-Li1-O4 | 99 | 99 | 0.00 |
| O2-Li1-O5 | 104 | 109 | 4.59 |
| O3-Li1-O4 | 134 | 117 | −14.53 |
| O3-Li1-O5 | 107 | 102 | −4.90 |
| O4-Li1-O5 | 112 | 135 | 17.04 |

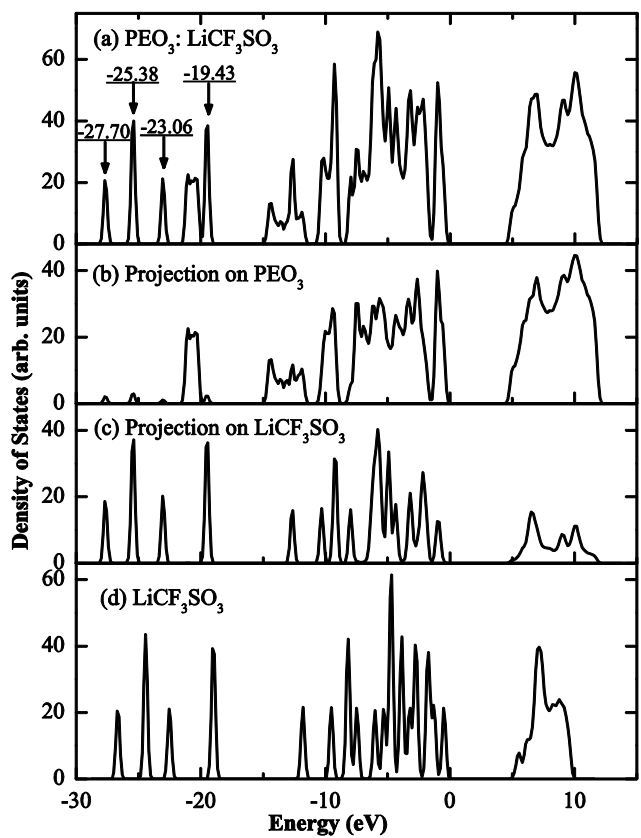

Fig. 4. Calculated density of states (DOS) for crystalline PEO$_3$:LiCF$_3$SO$_3$: (a) total DOS, (b) projected DOS on PEO$_3$, and (c) projected DOS on LiCF$_3$SO$_3$. As a comparison, density of state for isolated LiCF$_3$SO$_3$ is also shown in (d). The valance band maximum is located at 0 eV.

Another important conclusion from the DOS curves can be made about the hybridization of states due to $LiCF_3SO_3$ and $PEO_3$ throughout the entire energy range. For the lower energy range in Fig. 4(a) (lower than −15 eV), five peaks are clearly observed. While the contribution from $PEO_3$ is small, four of these five peaks are composed of states from both $LiCF_3SO_3$ and $PEO_3$, as seen in Figs. 4(b) and 4(c). Similarly, in the upper range, *i.e.*, between −15 eV and the valance band maximum, the $LiCF_3SO_3$ states are mixed with $PEO_3$ band. In fact, almost all the energy levels of isolated $LiCF_3SO_3$ shown in Fig. 4(d) are observed to shift down in energy after it is incorporated into $PEO_3$ [Fig. 4(c)]. For example, the four peaks of isolated $LiCF_3SO_3$ in the lower energy range [Fig. 4(d)] are shifted down by 0.99 eV, 0.95 eV, 0.51 eV, and 0.38 eV, respectively, to the locations of −27.70 eV, −25.38 eV, −23.06 eV, and −19.43 eV. All of these observations suggest bonding between $LiCF_3SO_3$ and $PEO_3$.

Table 5. Charge of each atom carries (in units of |e|) determined with Bader analysis and the charge distribution of the DFT calculation. Not shown are the charges of the 12 hydrogen atoms, which have a range of 0.05-0.12.

| $LiCF_3SO_3$ | | PEO | |
|---|---|---|---|
| Atom | Charge | Atom | Charge |
| Li | 0.88 | O1 | −1.08 |
| C7 | 1.63 | O2 | −1.05 |
| F1 | −0.61 | O3 | −1.06 |
| F2 | −0.61 | C1 | 0.40 |
| F3 | −0.61 | C2 | 0.36 |
| S | 3.26 | C3 | 0.39 |
| O4 | −1.34 | C4 | 0.43 |
| O6 | −1.31 | C5 | 0.39 |
| O7 | −1.34 | C6 | 0.35 |

Finally, as shown in Table 5, we present results of Bader charge analysis that provide a representation of charge transfers between atoms [31,32]. Charge analysis can also provide information about bonding characteristics [33-37]. As it can be seen from Table 5, Li carries a positive charge of 0.88, indicating that lithium in $LiCF_3SO_3$ is indeed a cation, as expected. The charge of the Li valance electron essentially transfers to the three oxygen atoms in $LiCF_3SO_3$, resulting in ionic bonding between $Li^+$ and $CF_3SO_3^-$. The three oxygen atoms also acquire negative charge from the sulfur atom. Similarly, the oxygen atoms in PEO are negatively charged with the charge transferred from the carbon and hydrogen atoms in PEO. There is also a significant charge transfer from the carbon atom in $LiCF_3SO_3$ to the nearby fluorine atoms.

Therefore, each of the C-O bonds in PEO, the S-O bonds in $LiCF_3SO_3$, and the C-F bonds in $LiCF_3SO_3$ has a mixed nature of both ionic bonding and covalent bonding. The C-S bond in $LiCF_3SO_3$ is, however, due entirely to covalent bonding as both the C and S atoms are positively charged, resulting in repulsive ionic interaction. Overall, both $Li^+$ and $CF_3SO_3^-$ have ionic interactions with PEO as each atom in PEO is charged, consistent with the DOS calculations.

## 4. Conclusion

In summary, *ab initio* density-functional calculations were performed to study the atomic and electronic structures of crystalline $PEO_3$: $LiCF_3SO_3$. The calculated results are shown to be consistent with the available experimental observations. In particular, the optimized lattice parameters are within ~3.0 % of the experimental values. The optimized geometry of crystalline $PEO_3$: $LiCF_3SO_3$ is also comparable with that determined by experiments. In addition, electronic structure calculations show the bonding between $LiCF_3SO_3$ and $PEO_3$ and the ionic conduction feature of crystalline $PEO_3$: $LiCF_3SO_3$.

Acknowledgements

This work was supported in part by NASA Grant No. NNX13AN01A and by Tulsa Institute of Alternative Energy and Tulsa Institute of Nanotechnology. The research used resources of the Extreme Science and Engineering Discovery Environment, the National Energy Research Scientific Computing Center, and the Tandy Supercomputing Center.